# Workflow description to dynamically model β-arrestin signaling networks

**Romain Yvinec*[1], Mohammed Akli Ayoub[1,2], Francesco De Pascali[1], Pascale Crépieux[1], Eric Reiter[1], and Anne Poupon[1]**

[1] PRC, INRA, CNRS, Université François Rabelais-Tours, 37380 Nouzilly, France.
[2] Biology Department, College of Science, United Arab Emirates University, PO Box 15551, Al Ain, United Arab Emirates

*__Corresponding author:__
Romain Yvinec, UMR PRC, F-37380 Nouzilly, France.
Phone : + 33 2 47 42 75 05 ; Fax : + 33 2 47 42 77 43 ; e-mail : romain.yvinec@inra.fr

**Abstract :** Dynamic models of signaling networks allow the formulation of hypotheses on the topology and kinetic rate laws characterizing a given molecular network, in-depth exploration and confrontation with kinetic biological data. Despite its standardization, dynamic modeling of signaling networks still requires successive technical steps that need to be carefully performed. Here, we detail these steps by going through the mathematical and statistical framework. We explain how it can be applied to the understanding of β-arrestin-dependent signaling networks. We illustrate our methodology through the modeling of β-arrestin recruitment kinetics at the Follicle Stimulating Hormone (FSH) receptor supported by in-house Bioluminescence Resonance Energy Transfer (BRET) data.

# 1. Introduction

β-arrestins have long been recognized to play a major role in G protein-coupled receptor (GPCR) desensitization and internalization. They are also known to transmit signals of agonist-activated GPCRs to promote G protein-independent signaling (1-3). β-arrestin-mediated signaling has been reported to impact gene expression, transcription (4) and translation (5).

β-arrestins serve as multiprotein scaffolds, connecting many partners (6), defining different pathways that act at different time-and-spatial scales (7-8). It is important to better understand β-arrestin signaling networks since it could lead towards the development of strategies, including the development of biased agonists (e.g. discriminating between G protein-dependent and β-arrestin-dependent signaling), with promising therapeutic implications (8-14).

The complexity of β-arrestin-mediated cellular regulation motivates the use of mathematical models that take into account its dynamic (*see* **Note** 4.1). Dynamic biochemical reaction network models, presented in this chapter, allow formulation of hypotheses on molecular networks linked to β-arrestins, by explicitly integrating detailed knowledge on signal transduction. The first impact of such framework is to reveal potential alternatives or conflicting signaling mechanisms that have been hypothesized in the literature. Subsequently, iterative confrontations of model and data facilitate hypotheses rejection or acceptance. Finally, this methodology can be used to predict the behavior of unobserved quantities : concentration of molecules, parameter values, etc. For further information, readers can refer to reviews and papers dealing with applications of GPCR signaling modeling (15-19).

At the very beginning of the modeling pipeline, one has to gather existing biological knowledge from the literature. In light of the available information, and according to the specific biological questions that are being addressed, the modeler decides the level of precision of the dynamic model. Hence, which molecules of a particular signaling sub-network are taken into account and how the reactions between molecules should be represented will drive the choice of a particular mathematical framework. The nature and amount of available data to be confronted with the model outputs is also of primary importance. We would like to emphasize that the use of time series experimental data, which is becoming more and more favoured due to the development of Bioluminescence Resonance Energy Transfer (BRET) or Fluorescence resonance energy transfer (FRET) technologies (20), is clearly a major improvement for quantitative dynamic modeling (17-19). Using perturbation experiments, which allow the observation of the system under different biological conditions, also clearly improves model selection.

Once a model has been defined, the modeler uses various tools to perform model analysis and numerical simulations, in order to gain knowledge on the behavior of the model outputs, but also to assess the impact of the model inputs and parameters (initial concentrations, reaction rates, etc.). Next, using an appropriate statistical description of the data, the modeler performs data fitting, with the aid of optimization algorithms, in order to assess the validity of the model. Parameter identifiability analysis, model selection and model predictions are major steps that provide key additional insight on the model's robustness and ability to accurately capture biological mechanisms. Of course, iterative model refinements and validation experiments are necessary to cement biological findings. We provide a graphical outline of the workflow in Figure 1. For specific applications of various modeling strategies, readers can refer to useful published reviews (15,21).

## 2. Materials

### 2.1 Software packages

We list below a few standard software systems we routinely use while modeling dynamic reaction systems.

1. To represent a reaction network diagram, we use Cell Designer (22) running on a standard laptop. See example in Figure 4.

2. To run model simulations, explore model behavior, and run some data fitting optimization, we use Copasi (23) running on a standard laptop.

3. To perform extensive data fitting optimizations, and parameter identifiability identification analysis, we use Data2Dynamics (or D2D), see (24), running on a standard laptop, with MATLAB (R2010 or newer). The results on our example presented in this chapter were obtained with D2D, see Figures 5, 6 and 7.

4. To perform extensive data fitting optimizations using global optimization algorithms , we use an in-house software HYPE (in preparation).

### 2.2 Plasmids

1. pcDNA3.1-hFSHR-Rluc8 (from A. Hanyaloglu, Imperial College, London, UK)

2. pcDNA3.1-yPET-ß-arrestin 2 (from M.G. Scott, Institut Cochin, Paris, France)

### 2.3 Cell culture and transfection

1. Human embryonic kidney (HEK293-FT) cells (Thermo Fisher Scientific, Massachusetts, USA)

2. Cell culture dishes (Corning Life Sciences, Tweksbury, MA, USA)

3. DMEM supplemented with 10% v/v fetal bovine serum, 4.5 g/l glucose, 100 U/ml penicillin, 0.1 mg/ml streptomycin, and 1 mM glutamine (all from Invitrogen, Carlsbad, CA, USA)

4. Trypsin-EDTA solution (Invitrogen, Carlsbad, CA, USA)

5. Phosphate buffered saline 1X without calcium and without magnesium (Invitrogen, Carlsbad, CA, USA) supplemented with HEPES 10 mM.

6. Transfection reagent: Metafectene PRO (Biontex, Munchen, Germany)

7. Multichannel pipette – 8 and 12 channels (Thermo Fisher Scientific, Massachusetts, USA)

### 2.4 BRET measurements

1. 96-well white microplates, flat-bottom, high binding, sterile (Greiner Bio-One, Courtaboeuf, France)

2. Coelenterazine h – 250 µg (Interchim, Montluçon, France)

3. Recombinant human Follicle Stimulating Hormone (hFSH) (FollMerck-Serono, Darmstadt, Germany)

4. Mithras$^2$ LB 943 plate reader (Berthold Technologies GmbH and Co. Wildbad, Germany) for BRET measurements.

## 3 Methods

The following modeling methods are illustrated with kinetic data obtained from cells displaying β-arrestin recruitment to the human Follicle Stimulating Hormone receptor (hFSHR) stimulated with seven different doses of hFSH. To assess this in live cells and in real-time we used BRET technology as previously reported (12,25) and described below.

### 3.1 Cell transfection

1. HEK293 cells are grown until 50% of confluence (10 x $10^6$ cells) in complete medium (DMEM supplemented with 10% v/v fetal bovine serum, 4.5 g/l glucose, 100 U/ml penicillin, 0.1 mg/ml streptomycin, and 1 mM glutamine) before their transient transfection with the plasmids encoding the different BRET sensors using Metafectene PRO following the manufacturer's protocol.

2. 5 µg of plasmid encoding FSHR-Rluc8 used as BRET donor and 5 µg of plasmid encoding yPET-ß-arrestin 2 used as BRET acceptor are mixed and resuspended in 1 ml of serum-free DMEM.

3. In parallel 50 µl of Metafectene PRO are resuspended in 1 ml of serum-free DMEM.

4. The two mixes (plasmids and Metafectene PRO) are first preincubated separately for 5 minutes at room temperature.

5. Then the plasmid solution is mixed with Metafectene PRO mix and the final solution (~2 ml) is incubated for 20 minutes at room temperature.

6. During the incubation of the transfection solution, cells are washed twice with PBS 1X, detached with 3 ml of Trypsin-EDTA, and resuspended to 10 x $10^6$ cells in a final volume of 18 ml of the complete medium.

7. After 20 minutes of incubation, the transfection solution (~2 ml) is mixed with 18 ml of cells in complete medium. This allows filling the whole 96-well plate with transfected cells using 200 µl per well.

8. Finally, 200 µl of plasmids + cells mix are distributed and seeded into each well of 96-well plate corresponding to $10^5$ cells/well and are cultured for forty-eight hours at 37ºC and 5% CO2 before BRET assays.

### 3.2 BRET measurements

1. Forty-eight hours after transfection, cells are first washed with 100 µl/well of PBS 1X.

2. Cells are then resuspended in 60 µl/well of PBS-HEPES 10 mM.

3. hFSH is prepared 5X in PBS-HEPES 10 mM at different doses for final concentrations of 0.0128, 0.064, 0.32, 1.6, 8, 40, and 200 nM.

4. Coelenterazine h (stock solution 1 mM) is first prepared 5X (25 μM) in PBS-HEPES 10 mM and kept on ice and protected from light.

5. For BRET measurements, to each 60 μl/well of cells in PBS-HEPES 10 mM 20 μl/well coelenterazine h 5X (5 μM final) is added, followed by injection of 20 μl/well of either PBS (vehicle-treated cells) or FSH (FSH-treated cells) prepared 5X in PBS-HEPES 10 mM at different doses as mentioned above.

7. Importantly, in order to avoid any kinetic delay between the different wells, a single well for each vehicle-treated and FSH-treated cells is read at a time and this is performed in independent quadriplicates.

### 3.3 Data pre-processing

1. The raw data are obtained as light emission at both 480±20 nm and 530±25 nm for each well including vehicle-treated (control) and FSH-treated cells. The Mikrowin 2010 software of the Mithras$^2$ LB 943 plate reader automatically calculates the conventional BRET ratio of 530±25 emission over 480±20 emission for each well and at every reading time point.

2. Since our focus is on FSH-promoted β-arrestin 2 recruitment, the data are then transformed to "Induced_BRET" values by subtracting the 530/480 ratio of vehicle-treated cells from its corresponding ratio of FSH-treated cells. This subtraction is done at each reading time point, as described previously (20,25,26). Our " Induced_BRET" data are shown in Figure 2.

3. To remove any replicate bias effect, we normalize them by a constant shift value so that all replicates for a given dose are as close as possible to each other (*see* **Note** 4.2). In Figure 3, we show the final data used in subsequent modeling steps.

### 3.4 Species and reaction list

1. Create and visualize a reaction network model using diagram editor software such as Cell Designer (see Figure 4). The following two steps are the key elements of a reaction network model (*see* **Note** 4.3 and **Note** 4.4), and may be performed by successive iterations while comparing the model to the data (*see* section 3.9). Competing hypotheses can be formulated at this step by dealing in parallel with different reaction network models ( *see* **Note** 4.6).

2. Define a list of molecules or species of interest to include in the model. Input variables, (partially or completely) controlled by the experimentalist, and output variables, measured by the experimentalist, should obviously be included. Intermediate molecules (enzymes, complex assemblies…) are taken into account according to the specific biological questions addressed ( *see* **Note** 4.5). In our example, we choose a list composed of five distinct species, representing various states of three molecules: (*FSH, FSHR, FSH-FSHR, β-arrestin, FSH-FSHR-β-arrestin*). In the modeling context, this list is known as the state variables.

3. Define a list of reactions between the species. These reactions should exactly define the stoichiometry of each species in each reaction. In our example, we include four reactions:

$$FSH + FSHR \rightarrow FSH\text{-}FSHR$$

$$FSH\text{-}FSHR \rightarrow FSH + FSHR$$

$$\beta\text{-arrestin} + FSH\text{-}FSHR \rightarrow FSH\text{-}FSHR\text{-}\beta\text{-arrestin}$$

$$FSH\text{-}FSHR\text{-}\beta\text{-arrestin} \rightarrow \emptyset$$

Hence, we have chosen to ignore GRK-mediated phosphorylation steps, and the internalization process is only represented as a sink reaction for the *FSH-FSHR-β-arrestin* complex (*see* **Note** 4.7).

### 3.5 Experimental conditions and observed variables

1. Enumerate the experimental conditions and the measured data that will be compared to the kinetic model. In our example, we have 7 conditions, corresponding to 7 different doses for the input concentrations of *FSH*. In each of these experimental conditions, we measured a BRET signal, which is transformed by a pre-processing step as an *Induced_BRET* signal (*see* sections 3.2 and 3.3).

2. Define the functions linking species of the model and measured data, including possible noise terms, coming from error measurements (*see* section 3.9 and **Note** 4.3). In our example, we choose the relation

*Induced_BRET = kf * FSH-FSHR-β-arrestin + ε,*

where $\varepsilon$ is the noise, defined as a Gaussian variable of mean 0 and unknown variance $\sigma^2$, and *kf* is an unknown proportional constant.

### 3.6 Parameter list

1. Once the reaction network and measured variables have been fully specified, we define the complete parameter set used in the model. For this, we assign to each state variable (i.e. a species of the reaction network), a parameter that specifies its initial quantity (its value at the starting experimental time, e.g. *t=0*). The initial concentrations are denoted in our example by

*(Init_FSH, Init_FSHR, Init_FSH-FSHR, Init_β-Arrestin, Init_FSH-FSHR-β-Arrestin)*.

Similarly, we assign to each reaction of the network a parameter that specifies its rate constant (*see* section 3.7). In our example, we denote the rate constants by (*kon, koff, k+, k-*) (*see* Figure 4)

2. For each experimental condition, one kinetic model (together with its observed variables) is constructed from the reaction network, its parameters, and the data/model relationship. Hence, each model has a set of initial quantities for the dynamical variables, a set of reaction rates for the reactions, and a set of measurement parameters for the observed variables. Note that in our example, we use the same reaction network for our 7 experimental conditions but it does not necessarily need to be the case in more general settings. Each of our 7 models has 11 parameters:

5 initial quantities : *(Init_FSH, Init_FSHR, Init_FSH-FSHR, Init_β-Arrestin, Init_FSH-FSHR-β-Arrestin)*$_{condition\ i}$, i=1...7.

4 kinetic rates: *(kon, koff, k+, k-)*$_{condition\ i}$, i=1...7.

2 measurement parameters: *(kf, σ²)*$_{condition\ i}$, i=1...7.

3. According to the experimental conditions, specify which parameters should have common values between any two conditions. The modeler thus may take into account possible perturbation experiments, where for instance some protein expression has been externally modified by the experimentalist (silencing RNA, over-expression, chemical inhibitor; *see* **Note** 4.8). In our example, because we deal with different doses of *FSH*, we take different values for the initial concentration of *FSH* for each condition, (*Init_FSH*)$_{condition\ i}$, but we assume that all remaining initial quantity parameters (*Init_FSH-FSHR, Init_FSH-FSHR-β-arrestin, Init_FSHR and Init_β-arrestin*), as well as all rate constants (*kon, koff, k+, k-*), and all measurement parameters (*kf, $\sigma^2$*) have common values across conditions (because the same transfection protocol was used for all experiments, *see* section 3.1).

4. In each experimental condition, specify whether the initial quantity for each state variable should be 0, a given known value, or a free parameter (that is, an unknown value that will be found during the data fitting procedure at section 3.9; *see* **Note** 4.9). In our example, we use the known input dose injected by the experimentalist for the initial quantity of *FSH*, (*Init_FSH*)$_{condition\ i}$, and we consider that *FSH-FSHR* and *FSH-FSHR-β-arrestin* are not present initially, so that *Init_FSH-FSHR=0* and *Init_FSH-FSHR-β-arrestin=0* (we do not take into account any constitutive activity). Finally, the last two initial quantities *Init_FSHR* and *Init_β-arrestin* are free parameters, and have identical values across the 7 experiments.

5. Repeat the previous step for rate constants and measurement parameters. In our example, the rate constants (*kon, koff, k+, k-*), and measurement parameters (*kf, $\sigma^2$*) are free parameters, and have identical values in the 7 experiments (*see* **Note** 4.10).

### 3.7 Dynamic rule

1. At this step, the modeler must choose a mathematical framework to simulate the time evolution of the molecules in the reaction network. There are essentially three choices. These choices must be made in compliance with the type of measured data. First, the modeler chooses to use discrete numbers or continuous concentrations to represent molecule abundance. Second, one chooses a stochastic or deterministic description (*see* **Note** 4.3). Third, the modeler specifies whether the spatial environment should be taken into account, and in this case, what mechanisms govern species displacements (Brownian motion in a continuous space, transitions between discrete compartments...). No spatial effect is taken into account when one assumes that the reaction network takes place in a single well-mixed compartment (*see* **Note** 4.11) Also, note that when the order of magnitude of the number of molecules is believed to be large, it is advised to adopt a continuous deterministic formulation rather than a discrete stochastic one, for computational efficiency (27) (it is also mathematically justified).

2. The above three choices lead the modeler to adopt one of these mathematical descriptions: petri nets, continuous time Markov chain (CTMC), stochastic differential equation (SDE), individual-based model (IBM), partial differential equation (PDE, reaction-diffusion equations) or ordinary differential equations (ODE). The reader can refer to (15,21) for further details. Here, for simplicity, and because it is the most common approach justified in standard cell culture experiments, we adopt the formalism of ordinary differential equations.

3. The modeler must finally choose how to represent reaction rate laws. In the ordinary differential equation context, we need to associate to each reaction a function that defines its velocity. For any reaction, its rate law is a function of the concentration of the reactants. The rate laws gives a non-negative number representing the speed of that reaction, which will impact the variation of both reactants and products. The higher this rate is, the faster the reactants are consumed and the products generated. The gold standard choice for the rate laws is given by the law of mass-action, which says that the rate of a reaction is proportional to the concentration of its reactants to the power of their stoichiometry (*see* **Note** 4.3). The proportional constant is the rate constant we assigned to each reaction in section 3.6.

### 3.8 Model analysis and simulation

1. As the reader may have noticed, dynamical models of biochemical reaction networks have a clear well-defined mathematical structure, deriving essentially from the species and reaction lists or topological model (*see* **Note** 4.3). As such, in practice, the modeler does not need to explicitly write down the equations governing the time evolution, neither to code a numerical scheme to simulate these equations. Indeed, most software (*see* materials section 2.1) in this field requires only as input the ingredients of the previous sections 3.4, 3.5, 3.6, 3.7.

2. According to the focus of the model and its objective, a first step to analyze it is given by the study of the long-time behavior: steady-state analysis, stability and bifurcation. The objective of such theoretical analysis is to be able to predict, according to the parameter values, whether the molecule abundances will converge towards steady values as time goes to infinity, or will disappear or keep increasing, or continuously oscillate. It is out of the scope of this methodological chapter to detail this theory (*see* **Note** 4.13).

3. When we are interested in transient behaviors, or when the reaction network is too complex to perform theoretical analysis, one needs to perform numerical simulation. The modeler thus needs to identify a given set of parameters (initial quantities, rate constants, measurement parameters) to run a time course of species abundance. Many software packages allow such numerical exploration, we can refer for instance to Cell Designer or Copasi (*see* materials section 2.1, s*ee* also **Note** 4.14).

4. A particularly useful way to numerically explore a model is to try to understand its parameter sensitivity. The objective is to explore the effect of changing a given parameter value on the observable variables. This notion is known as local parameter sensitivity, and it is defined mathematically as the first order derivative of the observable variables with respect to the parameters. Although it is relatively easy to calculate (it can be calculated together with the solution of the ODEs), it has the disadvantage to be a local notion, and hence will depend on the particular parameter set used, which is, at this step, still unknown! Copasi allows a systematic exploration of local parameter sensitivity. More sophisticated tools exist to define a notion of global sensitivity, and rely on sampling parameter sets on a given range of values (28).

### 3.9 Statistical model and data fitting

1. To define a statistical procedure for data fitting, the first step is to specify an observation model (or measurement model, *see* **Note** 4.3). The most common choice is an additive Gaussian noise of mean zero and constant variance, which are well suited for instance to the analysis of fluorescence measurement. We have used this choice, as already specified in the data/model relationship in section 3.5. In some applications, multiplicative log-normal noise may be a better choice (if the noise is asymmetric or higher for higher measured values). When we suspect the existence of outliers, Laplace distributed noise is shown to be robust (29). For other statistical models adapted to stochastic or PDE models, we refer to (30).

2. When the noise model has been defined, an objective function is defined through the notion of Likelihood ($L$): a function of the parameter that gives the conditional probability of observing the data, given the value of the model parameters. This conditional probability is directly linked to the choice of the error model defined at the previous step. In the statistical framework, we call the "frequentist approach" the process of finding the parameter sets that maximize the likelihood (*see* **Note** 4.15). However, for practical computing performance, one usually prefers to minimize $-log(L)$. For instance, the additive Gaussian noise model leads to minimize the standard weighted sum of squares. For more complex objective functions, we refer again to (29).

3. We then use an optimization algorithm to find possible parameter sets that minimize the objective function. Unfortunately, in concrete applications, the objective function is a nonlinear and non-convex function, for which there is no gold-standard method to find the global minimum. There is actually a plethora of choices to perform such tasks. We find that multi-start deterministic optimization based on random initial parameter values and gradient descent-like methods, like D2D (24), perform well. We also use global optimization methods combining two well-established techniques, namely genetic algorithm and CMA-ES (HYPE, in preparation). See (31,32) for other popular choices.

4. All numerical methods require initial parameter values to perform the optimization procedure, or a range to search in (or an initial probability distribution in the Bayesian context; *see* **Note** 4.15). Thus, the modeler specifies such values using order of magnitudes that are compatible with the biological knowledge. Conservative choices will look for sufficiently large parameter ranges (to make sure to find an optimal parameter set, at the price of increasing computational cost). In our example, we used wide initial ranges, of 10-20 logs (*see* **Note** 4.3).

5. Once enough optimization routines have been performed to presumably reach the optimal solution (*see* **Note** 4.16), the modeler must decide whether the solution is acceptable, or if the model should be modified to yield a better fit (and thus potentially back to section 3.4!). To make this decision, we use qualitative and quantitative information. Of course, visual inspection of the best fit is of importance. Final value of the objective function, if properly normalized (HYPE, in preparation), can be used, as well as convergence curves (24). The modeler may also use a residual test to assess the validity of both the kinetic model and the error model. Using a chi-squared or Kolmogorov-Smirnov test, we may test the adequacy of the difference between the data to the model with the probability distribution used for the error model. Finally, goodness of fit criteria may be used to assess for lack-of-fit or over-fitting.

In our example, we show the best fit together with residual test in Figure 5 and 6 respectively. As demonstrated by Figure 5, the model provides a satisfactory fit for the higher doses. The lower doses show a moderate discrepancy between model and data, as confirmed by residual plots and Kolmogorov-Smirnov test (see Figure 6). This bias could be removed by taking into account basal activity for instance.

### 3.10 Parameter identifiability, model selection

1. Theoretical (or structural) identifiability analysis aims at finding whether the optimization given by the data fitting problem, assuming perfect time-resolved data and no data error measurement, has a unique solution or not. A model is not structurally identifiable if two distinct parameter sets give exactly the same observable values. For simple to moderately complex models, several methods exist to assess structural (non-)identifiability (33-35). It is advised to perform such analysis before running computationally expensive numerical optimization procedures, as structural identifiability analysis may reveal over-parameterization (*see* **Note** 4.3) and can greatly simplify the numerical search. In such case, re-parameterization has to be considered (*see* **Note** 4.16). In our example, the structural identifiability analysis did not identify any over parameterization. For practical efficiency, we used time and concentration adimentionalization (*see* **Note** 4.3) in order to make most of the unknown parameters dimensionless. We used *Init_FSHR* and *k-* as reference for concentration and time units, respectively, and defined all other concentrations and kinetic rates relatively to both *Init_FSHR* and *k-* parameters.

2. A complementary useful approach is given by the practical identifiability. Based on local evaluation of the objective function, it allows to give confidence interval to the best parameter values (if it is identifiable), or to shed light on non-identifiability, given the data, by revealing parameter combination that leads to similar objective function value. Of importance, this method is not restricted to parameter values, but can be extended to any model output (36-38). In our example, practical identifiability analysis performed with D2D revealed that most parameters are not practically identifiable (Figure 7), suggesting that more data are needed (more doses, or different measurements like number of free receptor, or number of complex ligand-receptor for instance).

3. Often, the modeler is confronted to several alternative hypotheses for the biochemical reaction networks to represent a given set of data, either due to several competing biological hypotheses, or in the course of the iterative search of the model that best fits the data. Again, several criteria may be used to choose one or another model. A useful standard statistical test to choose the appropriate number of parameters in nested models is given by the likelihood ratio test.

Other quantities such as AIC (Akaike information criterion) or BIC (Bayesian information criterion) are often used to perform model selection (39).

4. Finally, once a satisfactory model has been selected, the modeler may use non-observable variables or different simulated experimental conditions to perform model predictions. To confirm these predictions, additional experiments are performed. In the case of inadequacy with the model predictions, the iterative confrontation of models to data continues (again, back to section 3.4).

# 4. Notes

4.1 We emphasize that this chapter is concerned with dynamic models, which have to be clearly distinguished with equilibrium models widely used in pharmacology (40). Dynamic models assess time-response relationships, while equilibrium models in pharmacology assess dose-response relationships.

4.2 Data processing: we use a simple linear model as follows. Denote by *Induced_BRET(t,j)* the induced BRET signal measured at time t and replicate j, we assume that *Induced_BRET(t,j) = N(m(t,j), $\eta^2$)*, where the mean is decomposed into two effects: *m(t,j) = m_t + s_j*. Thus, we model the induced BRET signal as a Gaussian variable of mean *m(t,j)* and constant variance $\eta^2$. The mean of replicate j is given by a time-dependent mean, shared by all four replicates, plus a replicate-dependent shift value given by s_j. Coefficient $\eta^2$, *m_t* and *s_j* are estimated using the standard lm function in R. Finally, to normalize the *Induced_BRET* signal, we subtract from each replicate its shift value *s_j*.

4.3 Additional mathematical notes are provided for the interested reader, at the weblink http://yvinec.perso.math.cnrs.fr/Publi/RY_17_betarrestin_review_math_notes.pdf

4.4 Reaction network construction: manual curation of the literature and biological knowledge should be taken into account in order to decide the relevance of intermediate partner molecules. Biomodel databases and pathway repositories may help to guide the modeler by starting from existing models. For a complete list, see Pathguide (http://www.pathguide.org). To look for protein interactions, good references to start with include BioGRID (https://thebiogrid.org/), STRING (https://string-db.org/), HIPPIE (http://cbdm-01.zdv.uni-mainz.de/~mschaefer/hippie/) or WikiPathways (http://www.wikipathways.org/index.php/WikiPathways). To look for existing models, see for instance Biomodels (https://www.ebi.ac.uk/biomodels-main/). Note that SBML (http://sbml.org/Main_Page) is a standard language to share signaling network models.

4.5 We designed an automatic procedure to complete existing signaling network models that could not explain given observations (17,41). Structural modeling of multi-protein complex assemblies (docking) can also help identifying partners and specific mechanisms (42,43).

4.6 We would like to emphasize the differences between an interaction or influence graph and a reaction network. In an influence graph, two molecules are linked if they interact with each other, but the mechanism is not specified (it could be a catalysis, or a complex formation, or other indirect mechanism…), in contrary to a reaction network. Often, one has a reasonable idea of the interaction graph, but several alternative reaction network models can be derived from it.

4.7 The use of the symbol ø does not necessarily mean that the species undergo a kind of degradation reaction, but rather that we do not wish to track in detail which subsequent reaction the species *FSH-FSHR-β-arrestin* undergoes. In a similar manner, spontaneous formation of a species *A*, out of an environment not modeled in detail, will be represented by a reaction ø → *A*.

Also, it is possible to model a reaction where the reactant is not consumed by also putting the species in the product side. For instance, transcription/translation processes are often modeled via reactions of the form *DNA → DNA + RNA, RNA → RNA + Protein*. A catalysis reaction may also be represented in this manner.

4.8 The advantage of using perturbation experiments is both to challenge the model (using a different range of behavior) and to increase the amount of data while not excessively increasing the number of parameters. We emphasize that the precise representation of the effect of a perturbation experiment is a modeling choice. As a rule of thumb, we choose to consider that all initial molecule concentrations may vary between two conditions when silencing RNA or over expression of a given molecule is used, because such modification usually occurs before the actual start of the experiment (a given external input stimulus is applied at time t=0 for instance). In contrary, when a specific chemical inhibitor is used, we modify only the initial concentration of its target, and consider that all other initial concentrations are not affected and thus remain the same. When the reaction rate is believed to be affected by a perturbation experiment (for instance if some enzyme not specifically modeled are shut down or over-expressed), we choose to modify the reaction rate by an extra multiplicative factor compared to the control experiment.

4.9 Some initial quantities can also be fixed using an equilibrium hypothesis. Depending on the experimental protocol, before the experiment actually starts (time of input stimulus for instance), we may consider that the variables are in equilibrium between each other. In such a case, initial quantity parameters can be expressed as a function of rate constants and other known initial quantities. For example, assuming that no FSH is present prior to injection by the experimentalist, this implies in our setting that Init_FSH-FSHR and Init_FSH-FSHR-β-arrestin both equal zero.

4.10 For complicated experimental design, we advise the use of a graphical network representation for each experimental condition, annotated by the particular parameter set used for this condition.

4.11 The single well-mixed compartment assumption of either continuous time Markov chain or ordinary differential equation formalism is clearly an approximation for live cell experiments. This approximation is often accepted in applications, and can be mathematically justified through homogenization techniques (44).

4.12 Other possible choices not discussed here are given by Michaelis-Menten or Hill kinetics (which need more than one parameter per reaction, but usually fewer species and reactions; see (45) ).

4.13 Exploiting the structure of biochemical reaction networks, a theory that goes back to the seminal works of (46-48) allows to derive sufficient conditions on the reaction network structure to guarantee specific long-time behavior, irrespectively of its parameter values. This theory has recently received lots of attention; see (49-53). Of practical importance for the modeler, one may for instance automatically check whether a reaction network satisfies some key properties to assess its long-time behavior. See for instance (54,55).

4.14 For more theoretical information on the numerical simulations, the reader can refer to (27, 28).

4.15 Alternative approaches not discussed here: the Bayesian approach search for the complete probability distribution of the parameters, according to the likelihood and prior information on parameter values. See (56) for more details.

4.16 There are many tricks in parameter estimation that are mostly dictated by computing performance and numerical robustness. a) It is often advised to adimentionalize the model. By doing so, only a few parameters carry physical units (mol, sec, etc.) while the remaining ones are defined as relative values and do not carry any units. b) In top of adimentionalization, re-parameterization can occasionally reveal some identifiable parameter combinations and is thus worth considering. c) Looking for parameter values in log scale allows the exploration of a larger state space in a computationally efficient manner, and is more adapted to biological data. d) Deciding when to stop a numerical search for a given model and a given parameterization is also non-trivial. Standard stopping criteria include the number of numerical evaluations of the objective function, the objective function value, the relative changes of the objective function between successive steps, and/or the relative changes of the parameters between successive steps.

**FIGURE Legends**

**Figure 1. Main steps of dynamical modeling of biochemical reaction networks.** On each red arrow, we indicate the corresponding section in the main text. Note that this diagram is not exhaustive, and more interactions between each step should be represented (in particular, as we explain in the main text, if the data fitting is not accurate enough, one needs to come back to either experimental design, reaction network construction or parameterization).

**Figure 2. Induced_BRET measurements for the recruitment of β-Arrestin at the FSHR after injection of a dose of FSH.** Each of the four replicates, for the seven FSH doses are plotted against time measurement (from t=0 to t=50). Dose 1 = 0,0128 nM;  Dose 2= 0,064 nM ; Dose 3 = 0,32 nM; Dose 4 = 1,6 nM ; Dose 5 = 8 nM ; Dose 6 = 40 nM ; Dose 7 = 200 nM.

**Figure 3. Normalized Induced_BRET measurements for the recruitment of β-Arrestin at the FSHR after injection of a dose of FSH.** Same as Figure 2 but with normalization within replicates to cancel any replicate bias (*see* section 3.3 and **Note** 4.2**)**.

**Figure 4. Reaction Network of a model for β-Arrestin recruitment at the FSHR**.  We include five  species, namely: FSH, FSHR, FSH-FSHR, Barr, FSH-FSHR-Barr, and four reactions (FSH + FSHR → FSH-FSHR, FSH-FSHR → FSH + FSHR, β-Arrestin + FSH-FSHR → FSH-FSHR-β-Arrestin, FSH-FSHR-β-Arrestin → ø) with rate constants respectively $K_{on}$, $K_{off}$, $K+$ and $K-$.

**Figure 5. Data fitting of the Induced_BRET.** The black line, plots the deterministic dynamic model (see reaction network in Figure 4) and the grey shaded area corresponds to the error measurement model (+/- 1.96 σ ). The color doted points are the normalized Induced_BRET measurements (same as Figure 3).

**Figure 6. Histograms of the residuals, obtained from the data fitting of Figure 5.** The p-values obtained from a Kolmogorov-Smirnov test (with mean 0 and variance 1 normal distribution) are respectively <1e-16, 7.3e-15, 3.2e-12, <1e-16, 0.006, 0.1 and 0.12 (for dose 1 to 7). Thus, normality test is accepted for the higher doses 6 and 7, and the test confirmed the visual shift of the residual for the lower doses.

**Figure 7. Profile likelihood estimates for each parameter.** In each subplot, we show the likelihood value around the optimized parameter set (black curve), the optimal likelihood value obtained during the optimization (dashed blue line) and the likelihood threshold value giving the 95% confidence interval for each parameter. Parameter values are in log scale.

The figures are available at http://yvinec.perso.math.cnrs.fr/Publi/RYetal_17_workflow_betaarr_fig.pdf
The supplemental mathematical notes are available at
http://yvinec.perso.math.cnrs.fr/Publi/RY_17_betarrestin_review_math_notes.pdf